\documentclass[journal=langd5,manuscript=letter]{achemso}

\usepackage[version=3]{mhchem} 

\author{Sujit S. Datta}
\author{Ho Cheung Shum}
\author{David A. Weitz}
\email{weitz@seas.harvard.edu}
\affiliation[Harvard University]
{Department of Physics and SEAS, Harvard University, Cambridge MA 02138}

\title{Controlled Buckling and Crumpling of Nanoparticle-Coated Droplets}

\begin{document}

\begin{abstract}
We introduce a new experimental approach to study the structural transitions of large numbers of nanoparticle-coated droplets as their volume is reduced. We use an emulsion system where the dispersed phase is slightly soluble in the continuous phase. By adding a fixed amount of unsaturated continuous phase, the volume of the droplets can be controllably reduced, causing them to buckle or crumple, thereby becoming non-spherical. The resultant morphologies depend both on the extent of volume reduction and the average droplet size. The buckling and crumpling behavior implies that the droplet surfaces are solid. \end{abstract}

\section{Introduction}
Suspensions of fluid droplets coated with solid colloidal particles are commonly known as ``Pickering'' emulsions. Particles with suitable surface chemistries can adsorb at the droplet surfaces strongly, with an energy of up to hundreds of $k_{B}T$. As a result, Pickering emulsion droplets can be exceptionally stable against coalescence and Ostwald ripening, making them useful for a wide variety of applications in which their interfacial properties play a key role. For example, Pickering emulsion droplets are promising candidates for encapsulating, delivering, or controllably releasing poorly soluble drugs \cite{Prestidge}. They have also been used as templates for fabricating hierarchical porous materials for catalysts or filters \cite{Gauckler}. 

	The stability of Pickering emulsion droplets is due to the dense packing of a strongly bound layer of colloidal particles at the fluid-fluid interface \cite{Cervantes, Langevin}; this layer is hypothesized to behave as a two-dimensional solid \cite{kralchevsky, vella, stratford, arditty, madivala}. However, experimental approaches to verifying this suggestion have been limited to {\it flat} fluid interfaces \cite{vella, aveyard1, aveyard2, zang, cicuta}. The solid-like nature of particle-coated surfaces is also often used to describe the curved interfaces of fluid droplets; for example, this assumption is essential to the interpretation of rheological measurements of densely packed Pickering emulsions \cite{arditty}. Nevertheless, direct experimental evidence of the solid-like nature of the surface of particle-coated droplets remains lacking.

	An unambiguous way to confirm the solid-like behavior of a thin shell is to observe buckling when it is strongly deformed. For example, as the volume of a thin elastic shell is reduced, compressive stresses build up on its surface, causing localized mechanical instabilities such as sharp points and bends to develop \cite{witten}. The proliferation of these localized instabilities results in a shell with a crumpled morphology. The crumpling of fluid-filled shells has been achieved experimentally either by uncontrollably dissolving or by evaporating the interior fluid for several porous capsules \cite{quilliet, tsapis, colver, studart}, or by using an externally imposed osmotic pressure \cite{gao, shum}. Few such studies exist for Pickering emulsions \cite{fuller}. A method to controllably induce compressive stresses at Pickering emulsion droplet surfaces is thus highly desirable to directly test their mechanical properties.

	In this Letter, we study the structural transitions of large numbers of nanoparticle-coated droplets as fluid is controllably pumped out of the droplet interiors. We find that a significant fraction of droplets buckle or crumple upon volume reduction, confirming the hypothesis that their interfaces behave like solids. The number of non-spherical droplets as well as the resultant droplet morphology is highly dependent on the amount of volume reduction and the average size of the droplets. All of the morphologies we observe are stable over a period of at least several hours. Many of these are strikingly similar to structures observed or predicted for buckled thin continuum elastic shells. The technique presented here provides a new and straightforward way to study the deformation behavior of thin fluid-filled granular shells.

\section{Results and Discussion}
	We use hydrophilic silica nanoparticles that are coated with a diffuse layer of alkane, rendering them partially hydrophobic and partially hydrophilic. Thus, they are wetted by both dispersed and continuous phases, and the three-phase contact angle characterizing the placement of the nanoparticles at the interface between the two phases is close to 90$^\circ$. As a result, once a nanoparticle is mechanically driven to this interface during the emulsification process, it is confined within a potential well of depth as large as hundreds of $k_{B}T$ \cite{pieranski}. This makes the emulsions used in this work highly stable for at least several weeks.
	
	Due to the strong attachment of the nanoparticles to the droplet surfaces, the average droplet size is controlled by the concentration of nanoparticles used to prepare the samples \cite{wiley}. The average droplet diameter $d$ is related to the volume concentration $c_{p}$ of nanoparticles with respect to the dispersed phase by $d = 8\phi_{A}na_{p}/c_{p}$, where $\phi_{A}$ is the fraction of the droplet surfaces that is covered with nanoparticles, $n$ is the average number of nanoparticle layers covering the droplets, and $a_{p}$ is the nanoparticle radius. This equation fits our data very well over a wide range, as shown in Figure 1. Using this fit, we estimate that the droplet surfaces are coated with a monolayer of nanoparticles with surface coverage $\phi_{A} = 80 \pm 2$\%.	
	
	A closed solid shell with a fixed surface area must deform as its volume is reduced and compressive stresses develop in the shell wall. These stresses are typically relieved through buckling events, in which the shell is strongly deformed only at points and lines on its surface \cite{witten}. Motivated by this, we develop a technique to study the mechanical properties of our Pickering emulsion droplet surfaces by reducing their average volume in a controlled manner. To achieve volume reduction, we use a dispersed phase whose solubility in the continuous phase is measured to be $\sim0.17$ vol\%. Thus mixing a Pickering emulsion with unsaturated continuous phase causes a finite amount of the dispersed phase to be pumped out of the droplets until the solubility limit of the continuous phase is reached. We calculate the average decrease in droplet volume $\Delta V$ relative to the original droplet volume $V_{0}$ to be $\Delta V/V_{0}\approx0.0017V_{add}/\phi V_{s}$, where $\phi$ is the emulsion volume fraction, $V_{s}$ is the initial sample volume, and $V_{add}$ is the volume of additional unsaturated continuous phase. For each value of $V_{add}$, we use optical microscopy to quantify the fraction of droplets that were clearly buckled or crumpled, analyzing a total of 3759 droplets. The pumping process presented here is analogous to applying a pressure difference across the particle-coated droplet surface. Similar to studies of osmotically-induced buckling of hollow capsules \cite{gao}, our technique involves the addition of a chemical species that forces solvent to be expelled from the droplet interiors. A key advantage of this approach is that it enables the simultaneous investigation of the buckling of large numbers of Pickering emulsion droplets of varying sizes, yielding statistics of this process unlike approaches using pendant drop tensiometry \cite{asekomhe} or a syringe \cite{fuller} to shrink individual macroscopic droplets.
	
	To study the mechanical nature of the particle-coated droplet surfaces, we observe their morphologies as their volume is reduced. Strikingly, a considerable fraction of the droplets become non-spherical due to buckling or crumpling (Figure 2), unambiguously demonstrating that their surfaces are solid. As $V_{add}$ and hence the degree of pumping is increased, more droplets are deformed, as shown in Figure 3. To further understand the properties of the droplet surfaces, we explicitly consider the microscopic interactions between the nanoparticles (Supporting Information). In the bulk continuous phase, the nanoparticles possess very weak attractions, whereas in the bulk dispersed phase the nanoparticles are much more strongly attractive due to the interpenetration of the alkyl chains coating the colloidal particles \cite{raghavan}, with attraction energy $\leq10^{4}k_{B}T$. Furthermore, nanoparticles confined to the fluid-fluid interface interact via capillary attractions \cite{kudrolli} of magnitude $\sim500k_{B}T$. Thus the densely-packed surfaces of the Pickering emulsion droplets are likely to be characterized by a non-zero two-dimensional shear modulus $G'_{2D}\approx40-700$mN/m either immediately after emulsification or upon slight droplet volume reduction (Supporting Information). We also use interfacial rheology to directly measure $G'_{2D}$ for the flat fluid-fluid interface saturated with an excess of nanoparticles; this shows that the layer of nanoparticles has mechanical properties similar to a soft glassy material \cite{cicuta} and has a magnitude consistent with our theoretical estimate (Supporting Information). Furthermore, because our nanoparticles are highly polydisperse, their structure at a droplet surface is likely to be amorphous, in contrast to systems with monodisperse particles in which crystalline defects play a significant role \cite{bausch, fortuna}. Thus, the droplet surfaces can be thought of as thin, porous, solid granular shells. The approach presented here is a powerful means to study volume-controlled morphological transitions of fluid-filled granular shells in a well-defined manner, by subjecting large numbers of droplets to the same degree of pumping.
	
	Unlike the case of a sheet, deforming a thin elastic shell requires it to both bend and stretch \cite{jellett}; thus, the buckling and crumpling behavior of such shells is typically understood by considering the energies associated with both of these deformation modes. Because stretching deformations require more energy than bending \cite{pogorelov}, they tend to be focused at points and lines, leading to localized mechanical instabilities that develop as the shell is increasingly deformed. While the mechanical properties of densely packed particles at fluid-fluid interfaces is typically interpreted within the framework of continuum elasticity \cite{kralchevsky, vella}, it is still not clear that this approach is appropriate for particle-coated droplets \cite{monteux, leahy}. To explore the similarities between classical thin elastic shells and our particle-coated droplets, we study the size dependence of droplet buckling as well as the morphologies of buckled and crumpled droplets. 
	
	By considering the shell size dependence of the energy required to form a depression in an elastic spherical shell, an analysis based on classical elasticity theory predicts that larger droplets should buckle more easily than smaller droplets \cite{quilliet}. To test this, we prepare samples consisting of droplets of three different average sizes, and controllably reduce their average volume. Crucially, our data clearly show that larger droplets require a smaller relative decrease in their volume than do smaller droplets to crumple by the same fraction, as shown in Figure 3. This suggests that larger droplets crumple more easily, which is in agreement with the idea of having a thin shell of densely-packed colloids covering the fluid droplets.
	
	Our approach is a useful way to characterize the morphological transitions undergone by droplets at varying degrees of pumping. For weak pumping, the droplets are often non-spherical, forming polyhedra with multiple flat facets separated by straight edges (Figure 4A). Similar shapes have been achieved previously for particle-coated bubbles or droplets that have coalesced or that have forcibly undergone plastic deformation \cite{stonenature}. In our system, however, it is likely that such shapes arise from the presence of localized defects in the granular shell. These can mediate the structural transitions of the shell due to volume pumping in a manner similar to that in which fivefold disclinations can lead to faceting of crystalline granular shells \cite{lidmar}. For intermediate pumping, droplets develop dimples or corrugations of size $\sim1-10\mu$m on their surfaces (Figure 4B). These are reminiscent of post-buckling structures for thin shells under pressure \cite{hutchinson} or core-shell structures whose shells have experienced significant compressive stresses \cite{gao, tsapis}. As pumping increases, droplets have more crumpled morphologies with increasing numbers of dimples developing; this is consistent with the notion that crumpling may be thought of as a sequential buckling process \cite{witten}. For stronger pumping, dimples grow into cusps and ridges spanning a wide range of size scales ($\sim0.1-100\mu$m) joining points of maximal curvature (Figure 4C). Due to the fixed surface area of the shell, these cusps and ridges start to interact with each other as fluid is pumped out of the droplets, as demonstrated by the increasing numbers of longer and sharper cusps and ridges shown in Figure 4C. This suggests that in-plane stretching of the nanoparticle network at the droplet surface is energetically more costly than bending, as in the case of thin continuum sheets, and stress is localized heterogeneously on the droplet surface.
	
	These observations strongly support the hypothesis that a sufficiently densely packed layer of colloidal particles at a fluid-fluid interface acts like a solid. In particular, the nonuniform capillary stresses associated with the highly deformed morphologies adopted by our droplets (Figures 2 and 4) must be supported by localized stresses in the solid granular shell; otherwise, these would be pulled back to a spherical shape by the fluid-fluid interfacial tension \cite{stonenature, stratford, clegg, cheng}. Moreover, all observed morphologies are ``frozen-in''; they are stable over an observation time of hours. Furthermore, when subjected to a shear flow, buckled droplets undergo rigid-body rotation as expected for a solid shell (Movie S1). This is clearly different from the tank treading motion expected for a droplet with a fluid surface \cite{keller}, confirming that the particles form a solid layer at the droplet interfaces.
	
	In addition to the typical morphologies shown in Figure 4A-C, droplets exhibit a number of other buckled structures. These include multifaceted polyhedral indentations consisting of up to eight vertices (Figure 4D and Figure S1A-C); sharply-defined creases and ridges that meet at individual points (Figure 4E); and elongated shapes strongly reminiscent of folded monosulcate lily pollen grains \cite{katifori} (Figure 4F). The observation of such a variety of buckled structures suggests that the mechanical properties of the interfaces of the Pickering droplets are non-uniform. The morphologies observed are strikingly similar to those seen for thin continuum elastic shells possessing locally more compliant areas on their surface. For example, a spherical cap with a single point-like ``weak spot'' develops a faceted polyhedral indentation within its surface upon volume reduction, similar to the structure shown in Figure 4D \cite{vaziri, katifori}. In folded sheets and large crumpled shells, multiple point-like singularities connected by stretching ridges give rise to morphologies similar to that shown in Figure 4E \cite{witten, mora, chaieb, blair}. A ``weak spot'' extended along a line on the surface of a shell, such as at a grain boundary \cite{bausch}, gives rise to an elongated structure upon volume reduction as in the droplet shown in Figure 4F \cite{katifori}. These comparisons suggest that the observed non-uniform crumpling morphologies can be attributed to heterogeneities in the colloidal shell of the Pickering droplets that inevitably arise during sample preparation. The compliant areas in the surfaces of Pickering emulsions likely guide the development of deformations during volume reduction, giving rise to the rich diversity of post-buckling morphologies. Thus, by tuning the arrangements of particles at the droplet interfaces, the crumpling morphologies can potentially be controlled. 
	
	In conclusion, we present a straightforward and general method for buckling and crumpling statistically significant numbers of nanoparticle-coated droplets by pumping fluid out of the droplet interiors. Increasing numbers of droplets buckle and crumple as their average volume is reduced. The extent to which droplets are buckled or crumpled depends on the average droplet size, with larger droplets crumpling more easily than smaller droplets in agreement with ideas based on continuum elasticity theory. As they undergo volume reduction, droplets become increasingly non-spherical and develop progressively more dimples, suggesting that crumpling may be thought of as a sequential buckling process. For stronger pumping, cusps and sharply curved ridges develop and grow. The structures of our buckled Pickering emulsions are strikingly similar to the buckled shapes of thin continuum elastic shells. Our approach is a new means to realize stable non-spherical fluid droplets, and our observations show that the droplet surfaces are solid-like. The technique presented here offers a route to systematically study the morphological transitions of thin fluid-filled granular shells undergoing volume reduction and to explore the similarities between these and ``classical'' thin elastic shells. 

\section{Experimental Details}
\subsection{Materials}Ethylene glycol (anhydrous, 99.8\%, Sigma-Aldrich), chlorobenzene (CHROMASOLV, for HPLC, 99.9\%, Sigma-Aldrich), glycerol (ACS reagent, 99.5\%, Sigma-Aldrich), and hydrophobic silica nanoparticles (diameter $\approx15$nm$\pm±30$\%, Tol-ST, Nissan Chemical Inc.) were used as received. Some samples were made using ethylene glycol and chlorobenzene dried by tumbling with molecular sieves (3\AA, beads, 4-8 mesh, Sigma-Aldrich) and filtered with a 0.2$\mu$m filter. The samples prepared with the dried reagents showed no noticeable difference in the final results, as compared to samples prepared with reagents used as received. 

\subsection{Emulsion Preparation} Emulsions were produced at a volume fraction of $\sim$61\% by mixing known masses of ethylene glycol containing 1\% glycerol (to minimize Ostwald ripening) and chlorobenzene containing a suitable amount of nanoparticle suspension. These were mechanically agitated by intensive vortexing for 10-40 minutes. This procedure forms extremely stable ``water-in-oil'' emulsions with ethylene glycol/glycerol as the dispersed phase and chlorobenzene as the continuous phase. This suggests that the silica nanoparticles have a contact angle close to 90$^\circ$, but are more preferentially wet by chlorobenzene \cite{binksopp}. Furthermore, these samples have surprisingly low polydispersities ($\sim$30\%) given the inhomogeneous shear rates that arise during emulsification. 

\subsection{Determination of Partial Solubility} Ethylene glycol is partially soluble in chlorobenzene at a concentration of $\sim$1.5$\mu$L/g. This is determined in two ways. First, the conductivities of both unsaturated chlorobenzene ($\sigma_{unsat}$) and chlorobenzene that has been saturated with ethylene glycol ($\sigma_{sat}$) were measured, as well as the conductivity of ethylene glycol ($\sigma_{EG}$). The saturation concentration of ethylene glycol in chlorobenzene ($c_{sat}$) was then calculated by linearly interpolating between $\sigma_{unsat}$ and $\sigma_{EG}$ using the relation $\sigma_{sat}\approx\sigma_{unsat}+c_{sat}\sigma_{EG}$. Second, the saturation concentration was directly measured by mixing measured amounts of ethylene glycol with chlorobenzene until the resulting solution is no longer visibly homogenous. Values of saturation concentration of ethylene glycol in chlorobenzene from the two measurements are in good agreement.

\subsection{Volume Control Procedure} To controllably pump a known volume of fluid from inside the droplets, a measured quantity of emulsion is added to a suitable amount of unsaturated chlorobenzene. This is then tumbled gently for several days. 

\subsection{Characterization} Droplet morphologies were characterized using optical microscopy at a range of magnifications (10x-100x), using either a Leica SP5 confocal microscope operating in bright-field mode or a Nikon Eclipse TE2000-E. Samples were imaged in sealed glass capillaries or sealed imaging chambers made from polyether ether ketone (PEEK) spacers.

\newpage

\acknowledgement

It is a pleasure to acknowledge V. Manoharan for use of the Nikon Eclipse TE2000-E microscope; K. Ladavac for experimental assistance in the early stages of this work; J. Sprakel for assistance with interfacial rheology measurements; R. Guerra, J. R. Hutchinson, T. E. Kodger, K. Ladavac, L. Mahadevan, M. Mani, D. R. Nelson, J. Paulose, H. A. Stone, and V. Trappe for useful discussions; and the anonymous referees for useful suggestions. This work was supported by the NSF (DMR-10006546) and the Harvard MRSEC (DMR-0820484).

\subsection{Supporting Information Available}
Discussion of microscopic interactions between particles, Image (Figure S1) and Movie (MovieS1.avi). This material is available free of charge via the Internet at {\tt http://pubs.acs.org}.

\bibliography{dattashumweitz2010}

\begin{mcitethebibliography}{8}
\bibitem[Not()]{Note-1}
This is a note. The text will be moved the the references section. The title of
  the section will change to ``Notes and References''.\relax
\mciteBstWouldAddEndPunctfalse
\mciteSetBstMidEndSepPunct{\mcitedefaultmidpunct}
{}{\mcitedefaultseppunct}\relax
\EndOfBibitem
\end{mcitethebibliography}

\newpage
\subsection{Figure 1}
\begin{figure}
    \includegraphics[scale=0.55]{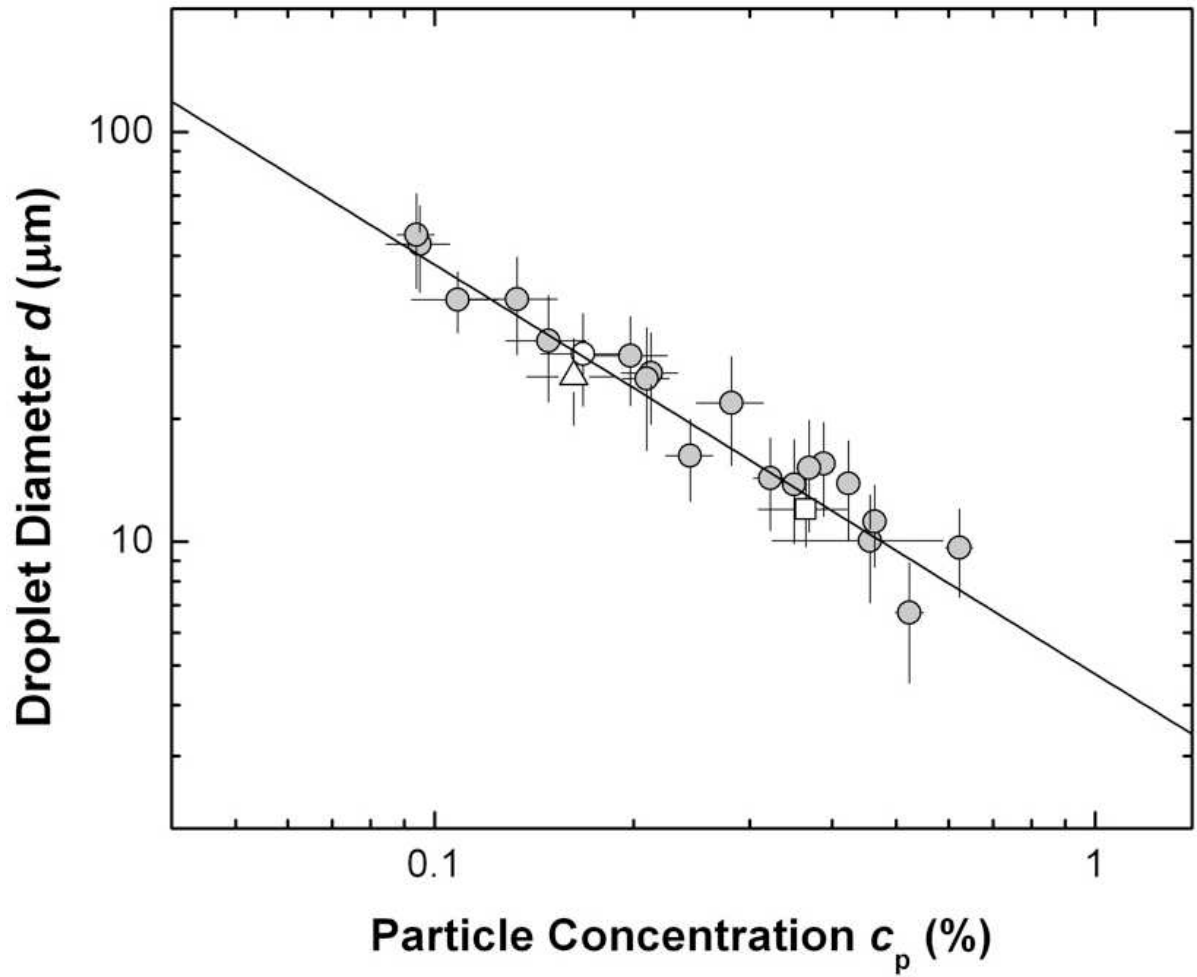}
  \caption{Average droplet diameter $d$ of Pickering emulsion prepared at different nanoparticle concentrations $c_{p}$. White square represents a sample made by adding nanoparticles to the sample represented by the white circle; white triangle represents a sample made by adding more dispersed phase to the sample represented by the white square. Vertical error bars indicate standard deviation of area-weighted size distribution, while horizontal error bars indicate estimated uncertainty in $c_{p}$. Solid line indicates inverse dependence of $d$ with $c_{p}$ ($d = 8\phi_{A}na_{p}/c_{p}$) as discussed in the text.}
  \label{Figure 1}
\end{figure}

\newpage
\subsection{Figure 2}
\begin{figure}
    \includegraphics[scale=0.5]{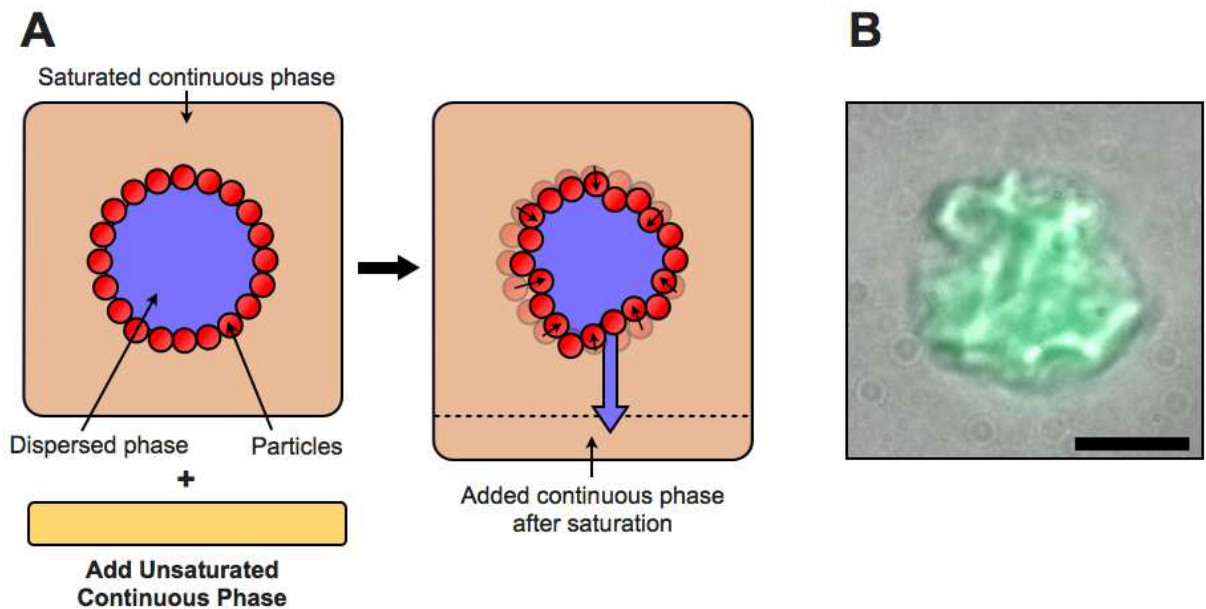}
  \caption{(A) Schematic illustrating the controlled reduction of droplet volume of Pickering emulsion droplets. (B) Optical micrograph of a crumpled droplet with fluorescently labeled dispersed phase (green overlaid); scale bar is 5$\mu$m.}
  \label{Figure 2}
\end{figure}

\newpage
\subsection{Figure 3}
\begin{figure}
    \includegraphics[scale=0.85]{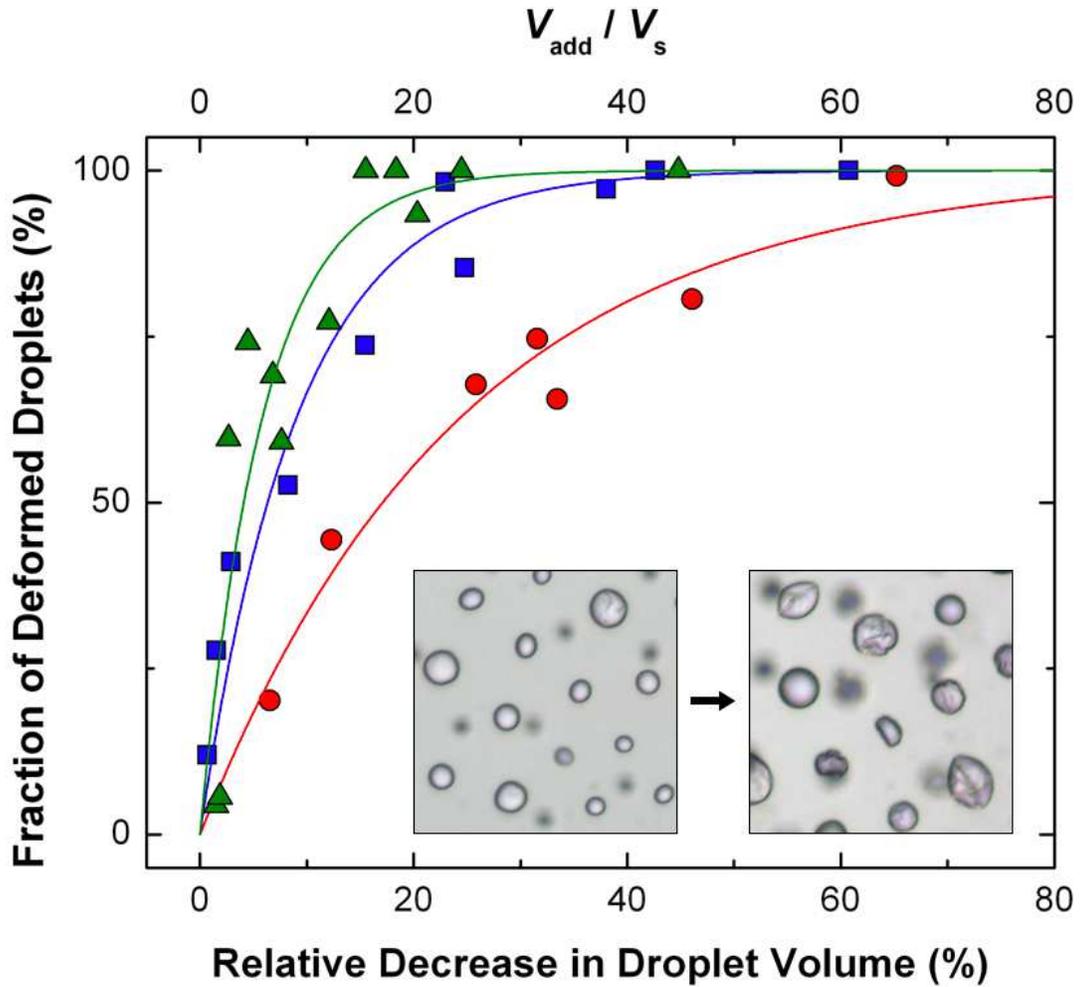}
  \caption{A plot of the fraction of buckled/crumpled droplets as a function of calculated relative change in droplet volume for samples of average droplet diameters $d=14.7\mu$m (red circles), $34.7\mu$m (blue squares), and $44.1\mu$m (green triangles). Solid lines are guides to the eye. Inset shows optical micrographs of two different samples: left has undergone weak pumping, right has undergone stronger pumping.}
  \label{Figure 3}
\end{figure}

\newpage
\subsection{Figure 4}
\begin{figure}
    \includegraphics[scale=0.45]{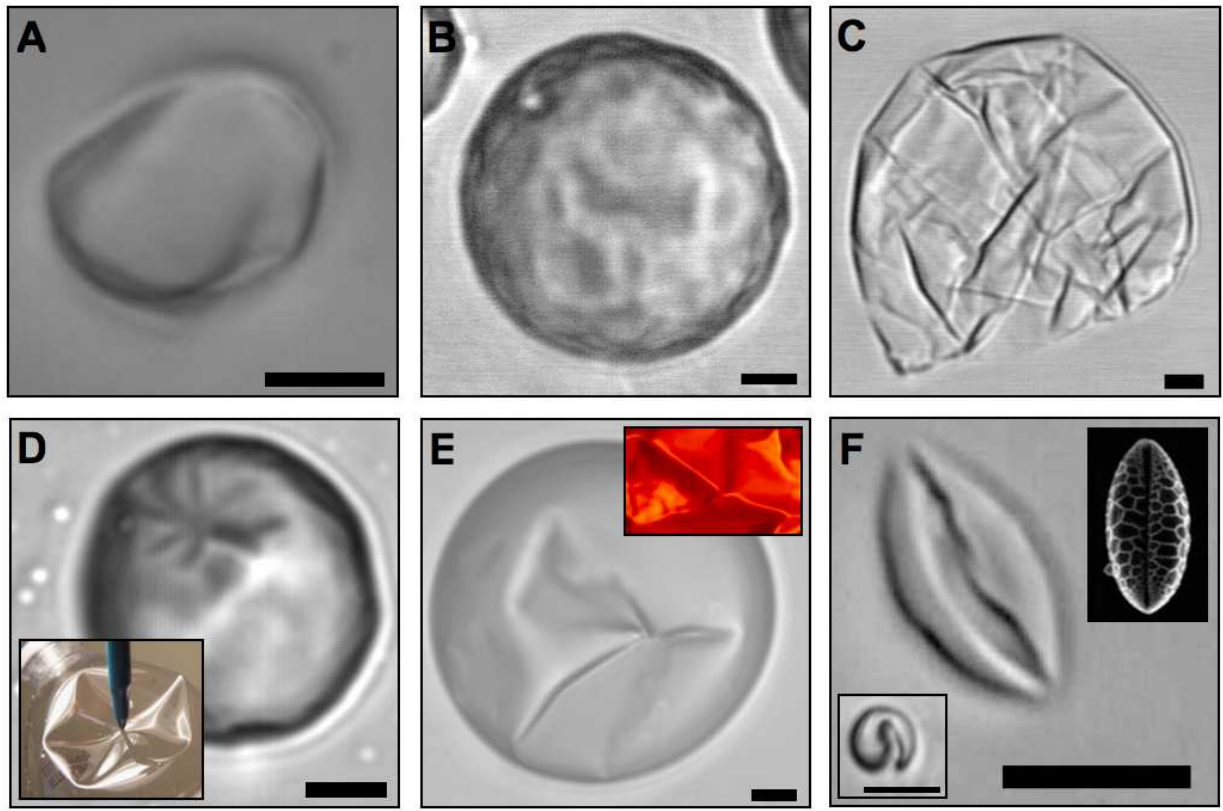}
  \caption{Optical micrographs of different buckled droplets. (A-C) show characteristic shapes at increasing levels of pumping, ranging from weak pumping (A), intermediate pumping (B), and strong pumping (C). Typical buckled structures also include multifaceted polyhedral indentations (D), stretching ridges, {\it d}-cones, and ``swallowtail'' folds (E), and elongated ``football'' structures (F). Bottom-left inset to (F) shows representative end-on view. All scale bars are 5$\mu$m. Colored insets show macroscopic analogues of buckled structures: point indentation of a bottle using a pen \cite{vaziri} (D), creases and ridges in crumpled paper \cite{blair} (E), and elongated dessicated pollen granule \cite{katifori} (F). }
  \label{Figure 4}
\end{figure}

\end{document}